\documentclass[11pt]{article}
\usepackage{times}
\usepackage{pgf,pgfarrows,pgfautomata,pgfheaps,pgfnodes,pgfshade}
\usepackage{graphicx}
\usepackage{epstopdf}
\usepackage{picins}
\usepackage{epsf}
\usepackage{subfigure}
\usepackage{amsmath}
\usepackage{amssymb}
\usepackage{amsfonts}
\usepackage{epsfig}

\def\g{{\cal G}}
\def\gup{{\cal G}^{+}}
\def\gdp{{\cal G}^{++}}
\def\G{{\cal G}^{+++}}

\begin{document}
\thispagestyle{empty}
\setcounter{page}{0}
\renewcommand{\theequation}{\thesection.\arabic{equation}}

{\hfill{\tt hep-th/0701165}}

{\hfill{ULB-TH/ 07-03}}

\vspace{1cm}

\begin{center} {\bf \large  Gravitational theories coupled to matter as invariant theories under Kac-Moody algebras\footnote{Talk presented at the 2nd RTN Workshop of the RTN project 'Constituents, Fundamental Forces and Symmetries of the Universe', Napoli, October 9-13, 2006.}}

\vspace{.5cm}

 Nassiba Tabti\footnote{F.R.I.A Researcher.}

\footnotesize \vspace{.5 cm}

{\em Service de Physique Th\'eorique et Math\'ematique,
Universit\'e Libre de Bruxelles,
\\ Campus Plaine C.P. 231\\ Boulevard du Triomphe, B-1050 Bruxelles,
Belgium}

\vspace{.2cm}

{\em The International Solvay Institutes, \\Campus Plaine
C.P. 231\\ Boulevard du Triomphe, B-1050 Bruxelles, Belgium}\\
\vspace{.2cm}

{\tt ntabti@ulb.ac.be}

\end{center}

\vspace {1.5cm}

\centerline{\bf Abstract} \noindent 
Many recent researches indicate that several gravitational $D$-dimensional theories suitably coupled  to some matter fields (including in particular pure gravity in $D$ dimensions, the low energy effective actions of the bosonic string and the bosonic sector of M-theory) would be characterized by infinite dimensional Kac-Moody algebras $\mathcal{G}^{++}$ and $\mathcal{G}^{+++}$. The possible existence of these extended symmetries motivates a development of a new description of gravitational theories based on these symmetries. The importance of Kac-Moody algebras and the link between the $\mathcal{G}^{+++}$-invariant theories and the uncompactified space-time covariant theories are discussed. 

\newpage
\baselineskip18pt

\setcounter{equation}{0}
 \addtocounter{footnote}{-1}
\section{Introduction}
The best candidate for the description and unification of all fundamental interactions is M-Theory in eleven dimensions.Very little is known for sure about M-Theory  but it is thought to
encompass all superstring theories and in particular, in low
energy limit, the eleven dimensional supergravity. Essential elements are lacking in the quest for a unified theory of quantized gravity and
matter. 
In this context, the study of hidden symmetries,
exhibited by dimensional reduction, would allow a better understanding
of the structure of the unified theory and could lead to a new formulation of gravitational interactions. These hopes have
been encouraged by some developments from recent years that certain
types of Kac-Moody algebras occur in several $D$-dimensional theories of gravity suitably coupled to dilatons and matter fields associated to $n$-forms, whose lagrangian is:
\begin{eqnarray}\label{actionmax}
\mathcal{L}_{D}= \sqrt{-g}\,\bigg( R -
\frac{1}{2}\sum_{u=1}^{q}\partial_{M}\Phi^{u} 
\partial^{M}\Phi^{u} -\sum_{n}\, \frac{1}{2n!}\, e^{\sum_{u}a_{n}^{u}\Phi^{u}}
F^{2}_{(n)}   \bigg) \, .
\end{eqnarray}
The possible existence of these extended symmetries motivates a development of an original formulation of gravitational theories based on these Kac-Moody algebras.
%Furthermore, it has been argued that a
%rank twenty-seven Kac-Moody algebra called $D_{24}^{+++}$ underlies
%the closed bosonic string in twenty-six dimensions and that even
%pure gravity in $D$ dimensions possesses a Kac-Moody algebra of rank
%$D$ $A_{D-3}^{+++}$ \cite{Lambert:2001gk}.\\
% Furthermore, it has been argued that a rank twenty-seven Kac-Moody algebra called $D_{24}^{+++}$ underlies the closed bosonic string in twenty-six dimensions and that even pure gravity in $D$ dimensions possesses a Kac-Moody algebra of rank $D$ $A_{D-3}^{+++}$ \cite{Lambert:2001gk}.
\section{Dimensional reduction and coset symmetries}
In this section, we will explain how Kac-Moody hidden symmetries are exhibited. Performing first a compactification on a torus $T^{D-3}$ of a $D$-dimensional gravitational theory given by Eq.(\ref{actionmax}), we find in $3$ dimensions a lagrangian containing only scalars coupled to gravity:
\begin{eqnarray}\label{lagred}
\mathcal{L}_{3}=
\sqrt{-g} \, \bigg(R - \frac{1}{2}\,
\partial_{\mu}\bar{\varphi} \cdot \partial^{\mu}\bar{\varphi} -
\frac{1}{2}\,\sum_{\bar{\alpha}}\, e^{\sqrt{2}\,\bar{\alpha}\cdot
\bar{\varphi}}\, \partial_{\mu} \chi_{\bar{\alpha}} \,
\partial^{\mu} \chi_{\bar{\alpha}} \bigg)\, .
\end{eqnarray}
One expects that the symmetry of this reduced lagrangian will be $GL(D-3,\mathbb{R})$ which is the symmetry group on the $D-3$ torus. But for some very specific theories, this symmetry is much larger. In fact, under certain conditions\footnote{A necessary condition that $\mathcal{L}_{3}$ is invariant under $\g$ and not only under $GL(D-3, \mathbb{R})$ is that the $\bar{\alpha}$ in Eq.(\ref{lagred}) are identified to positive roots of $\g$. },   the scalar part of the reduced lagrangian $\mathcal{L}_{3}$ can be identified to a coset lagrangian $\mathcal{L}_{\g/\mathcal{K}}$ invariant under transformations $\g/\mathcal{K}$ where $\g $ is a simple Lie group and $\mathcal{K}$ is the maximal compact subgroup of $\g$ \cite{Cremmer1979up,Marcus1983hb}. \\
For instance, upon dimensional reduction down to $3$ dimensions, the bosonic part of the $11$-dimensional supergravity whose lagrangian is written as $ \mathcal{L}_{11} = \sqrt{-g}\, ( R- \frac{1}{2.4!} \ 
 F_{\mu\nu\rho\sigma}F^{\mu \nu \rho \sigma}+  \ C.S. \ )$ exhibits the simple Lie group $\g=$ $E_8$. The Dynkin diagram of $E_8$ (see Fig.1) characterizes completely this simple Lie algebra. The vertices on the horizontal line define the gravity line. They represent simple roots related to fields coming from the metric $g_{\mu \nu}$. The vertex not belonging to the gravity line is related to a field resulting from the $4$-form $F_{\mu \nu \rho \sigma}$.\\

%\begin{eqnarray}
% \mathcal{L} = \sqrt{-g} \bigg( R- \frac{1}{2.4!} \ 
% F_{\mu\nu\rho\sigma}F^{\mu \nu \rho \sigma}+  \ C.S. \ \bigg)
%\end{eqnarray}
 
%\begin{center}
%\scalebox{.5}{
%\begin{picture}(180,60)

%\thicklines \multiput(-80,10)(40,0){7}{\circle{10}}
% lignes entre les vertex
%\multiput(-75,10)(40,0){6}{\line(1,0){30}}
%1 vertex du dessus
%\put(80,50){\circle*{10}}
%ligne vers le haut
%\put(80,16){\line(0,1){30}}% \put(0,-25){Fig.1 - \small .}
%\end{picture} 

%}
%\end{center}
%\vchcaption{Dynkin diagram of $E_{8}$}

It is interesting to consider the dimensional reduction beyond $3$ dimensions. That would give Dynkin diagrams of Kac-Moody algebras. Roughly speaking, Kac-Moody algebras are infinite dimensional generalization of finite dimensional Lie algebras $\g$.  It has been showed that the reduced theory to $2$ dimensions are connected to a infinite dimensional symmetry $\gup$ ( affine extension of $\g$) \cite{Nicolai1987vy} obtained by adding one root to the Dynkin diagram of $\g$. Motivated by the dimensional reduction, it has been argued that  the  Kac-Moody algebra $\gdp$ (overextension of $\g$) can play a role in the compactification to $1$ dimension \cite{Julia1982gx}. Finally, when all the dimensions are compactified, it is obvious to extend the algebra $\gdp$ to $\G$ (triple extension of $\g$) by adding a third vertex. Such $\G$ symmetries were first conjectured in the aforementioned cases 
\cite{West2001as, Lambert2001gk} and the extension to all $\G$ was proposed in \cite{Englert2003zs}. So this construction motivates the fact that the eleven dimensional supergravity could have the symmetry $\G=E_8^{+++}$ (see Fig.1).\\
 
 Similarly other well known theories would also have  symmetries under Kac-Moody algebras. Indeed the pure gravity in $D$ dimensions and the effective action of the bosonic string in $26$ dimensions would possess respectively a very-extended symmetry $A_{D-3}^{+++}$ and $D_{24}^{+++}$. More generally all simple maximally non-compact Lie groups $\g$ could be generated from the reduction down to $3$ dimensions of suitably chosen actions \cite{Cremmer1999du} and it was conjectured that these actions possess the very-extended Kac-Moody symmetries $\G$ \cite{Englert2003zs}. $\G$ algebras are defined by the Dynkin diagrams depicted in Fig.1, obtained from those of $\g$ by adding three nodes. % Indeed this symmetry could exist for the non-reduced theory in a complex manner and be easily exhibited by the dimensional reduction process. 
\begin{figure}[!h]
  \centering
{\scalebox{.41}
{\includegraphics{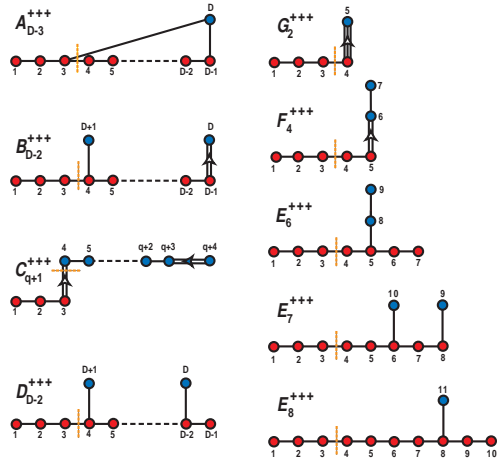}}}
 \caption { \small The
nodes labeled 1, 2, 3 define the Kac-Moody extensions $\G$ of the  Lie
algebras $\g$. The horizontal line
starting at 1 defines the `gravity line', which is the
Dynkin diagram  of a
$A_{D-1}$ sub-algebra.}
   \label{fig:2}
\end{figure}
\newpage
\section{$\G$- invariant action}
To make the Kac-Moody symmetries manifest, it is interesting to construct an action explicitly invariant under $\G$: $\mathcal{S}_{\G}$ \cite{Englert2003py}. This
action is defined in a reparametrisation invariant way  on a
world-line, a priori unrelated to space-time, in terms of an infinity of fields
$\varphi(\xi)$ where $\xi$ spans the world-line. The fields $\varphi(\xi)$
live in a coset space
$\G/\mathcal{K}^{+++}$ where the sub-algebra $\mathcal{K}^{+++}$ is invariant under a
`temporal involution' $\Omega$
 which ensures that the action is $SO(1,D-1)$ invariant. We can write an element of the coset by exponentiating the Borel sub-algebra  of $\G$ (which contains the Cartan and the positive root generators) as:
 \begin{eqnarray}\label{corep}
 \mathcal{V} (\xi)= exp^{\mathcal{B}^{a} \phi_a(\xi) }\, .
 \end{eqnarray}
  To each Borel generator $\mathcal{B}^a$, we associate a field $\phi_a(\xi)$. As there is an infinity of generators $\mathcal{B}^a$, there is an infinite number of fields $\phi_a(\xi)$. So we have to organize this summation on the infinity of fields in such a way that a recursive approach has a sense. In this context, we introduce a level decomposition with respect to the finite sub-algebra $A_{D-1}$  where $D$ is interpreted as the space-time dimension. Each $\G$ contains indeed a sub-algebra $GL(D)$ such that $SL(D)(=A_{D-1}) \subset GL(D) \subset
\G$.\\

 In the particular case of $E_8^{+++}$, the level $l$ counts the number of times the simple root $\alpha_{11}$ (not contained in the gravity line) appears in irreducible representation of  $A_{10}$ \cite{bi10}: $ \alpha = l \alpha_{11} + \sum_{i=1}^{10} a_i \alpha_i \, .$
Using this mechanism, we find a nice decomposition of the infinite number of generators. At each level we have a finite number of generators and fields associated to it.  Exploring  the decomposition of $E_8^{+++}$ into representations of $A_{10}$, the positive root generators at level 0, 1, 2 and 3 are respectively, $K^a_{\ b}$ ($GL(11)$ generators), $R^{a_1a_2a_3} $ , $R^{a_1a_2a_3a_4a_5a_6}$ and $R^{a_1a_2a_3a_4a_5a_6a_7a_8,b}$. The level $1$ and $2$ tensors are antisymmetric  and the level $3$ tensor is associated to a mixed Young tableau (with the constraint $R^{[a_1a_2a_3a_4a_5a_6a_7a_8,b]}=0$). The fields associated to these generators are respectively $h_a^b(\xi)$, $A_{a_1a_2a_3}(\xi)$,
$A_{a_1a_2a_3a_4a_5a_6}(\xi)$ and $A_{a_1a_2a_3a_4a_5a_6a_7a_8,b}(\xi)$. At higher levels ($> 3$), there is an infinite number of representations characterized by some Young tableaux. \\

With this decomposition, it is possible to rewrite the coset representative $\mathcal{V}(\xi)$ in Eq. (\ref{corep}) as
\begin{eqnarray}
\mathcal{V}(\xi)= exp\bigg(\underbrace{\!\sum_{a\geq b}h_{b}^{\
a}(\xi)K^{b}_{\ a}}_{Level\
0}\!\bigg)exp\bigg(\underbrace{\!\sum A_{a_1\, a_2 \,a_3}(\xi)\, R^{a_1\, a_2 \,a_3}+ \ldots }_{Level\ \geq\,1}\!\bigg)\, ,
\end{eqnarray}
where the first exponential contains only level zero operators and the second one the positive root generators of levels strictly greater than zero. Defining
\begin{eqnarray}
\frac{dv(\xi)}{d\xi}=\frac{d\mathcal{V}}{d\xi}\mathcal{V}^{-1} \ , \qquad 
\frac{d{\tilde{v}}(\xi)}{d\xi}=-\Omega _{1}\, \frac{dv(\xi)}{d\xi}=
\tilde{\mathcal{V}}^{-1}\frac{d\tilde{\mathcal{V}} }{d\xi}\ ,\qquad 
\mathcal{P}=\frac{1}{2}\, \bigg (\frac{dv}{d\xi}+\frac{d{\tilde{v}}}{d\xi} \bigg ) \, ,
\end{eqnarray}
where $\Omega_1$ is the temporal involution which allows identification of index $1$ to a time coordinate, one obtains in terms of the $\xi$-dependent fields, an action $\mathcal{S}_{\G}$ invariant under global $\G$ transformations, defined on the coset $\g/\mathcal{K}^{+++}$ :
\begin{eqnarray}\label{actionG}
\mathcal{S}_{\mathcal{G}^{+++}}=\int \, d\xi \,
\frac{1}{n(\xi)}\big<\mathcal{P} \mid \mathcal{P}
\big> \, ,
\end{eqnarray}
where $n(\xi)$ is an arbitrary lapse function ensuring reparametrisation
invariance on the world-line and '$< , >$'  is the $\G$ invariant bilinear form ensuring invariance of the action (\ref{actionG}) under the global transformations $\G$.\\

 The fundamental question is now : is there a link between the action invariant under the Kac-Moody algebras and the action of the space-time theory? In particular, considering our example, we would like to relate the action invariant under $E_8^{+++}$ and the $11$- dimensional supergravity. To do that, we must interpret the parameter $\xi$ which the fields of the $\sigma$-model depend on and find the significance of the infinity of fields $\phi_a(\xi)$. In this context, we are going to study the Kac-Moody algebra $\gdp$.
\section{$\gdp$- invariant actions}
To make connections between this new formalism and the covariant
space-time theories, it is interesting to analyze the several
actions invariant under overextended Kac-Moody algebra $\gdp$
(double extension of $\g$). The Dynkin diagram of $\gdp$ is obtained by deleting the root $\alpha_1$ in the Dynkin diagram of $\G$. The $\G$-invariant action
leads to two distinct theories: $\gdp_{C}$ and $\gdp_{B}$ obtained
both by a truncation of a infinity of $\G$-fields, putting to zero all the fields multiplying generators involving the deleted root $\alpha_1$. This truncation is realized consistently
with all equations of motion, i.e. it implies
that all the solutions of the equations of motions of
$\mathcal{S}_{\gdp}$ are also solutions of the equations of motion
of $\mathcal{S}_{\G}$  \cite{Englert2004ph}.
\subsection{$\gdp_{C}-$ invariant action}
The recent study of the properties of cosmological solutions in the vicinity of a space-like singularity revealed an overextended symmetry $\gdp_C$. The action $\mathcal{S}_{\gdp_C}$ restricted to a defined number of lowest levels is equal to the corresponding space-time theory in which the fields depend only on the time coordinate (see Table 1) \cite{Damour2002cu}. The parameter $\xi$ is defined as a time coordinate and this $\gdp_C$-theory carries a Euclidean signature in $D-1$ dimensions. 

\begin{table}[!h]
\centering
\scalebox{0.83}{
\begin{tabular}{cccc}
&\emph{Fields belonging  to $\mathcal{S}_{E_{8}^{++}}$}  & & \emph{Fields of supergravity depending on time} \\
&& &  \\
level 0& $g_{\hat{\mu} \hat{\nu}}({\color{blue}t})= fct(h_a^{\ b})$  &  $\leftrightsquigarrow $  & metric \\ 
level 1&  $A_{\hat{\mu}\hat{\nu} \hat{\rho}}({\color{blue}t})$ &  $\leftrightsquigarrow$  & 3-form electric potential \\
level 2&  $A_{\hat{\mu}_1...\hat{\mu}_6}({\color{blue}t})$ &  $\leftrightsquigarrow$  & 6-form magnetic potential (dual of the 3-form) \\ 
level 3&  $A_{\hat{\mu}_1...\hat{\mu}_8,\hat{\nu}}({\color{blue}t})$ &  $\leftrightsquigarrow$  & 'dual' of the metric
\end{tabular}
}
\caption{Link between the lowest level up to level $3$ of the cosmological $E_8^{++}$-invariant theory and the fields of the $11$-dimensional supergravity.}
\end{table}
\subsection{$\gdp_{B}$-invariant action}
The action $\mathcal{S}_{\G}$ contains another $\gdp$-invariant action $\mathcal{S}_{\gdp_{B}}$, obtained by performing the same consistent truncation as the one for $\gdp_{C}$, but now performed after a $\G$ Weyl transformation. As the Weyl transformations  modify the identification of the indices $1$ and $2$ which become respectively space and time, the resulting action $\mathcal{S}_{\gdp_B}$ is different from $\mathcal{S}_{\gdp_C}$. This gives an action $\mathcal{S}_{\gdp_B}$ with a Lorentz signature for the
metric and with the parameter $\xi$ identified to the missing
space coordinate instead of $t$. This $\gdp_B$-theory admits exact solutions identical to the ones of covariant Einstein and field equations describing intersecting extremal brane solutions smeared in all 
direction but one (see Table 2) \cite{Englert2003py,Englert2004it}.
\begin{table}[!h]
\centering
\scalebox{.83}{
\begin{tabular}{cccc} 
&\emph{Fields belonging  to $\mathcal{S}_{E_{8}^{++}}$}  & & \emph{Branes of M-theory} \\
&& &  \\
level 0& $g_{\hat{\mu} \hat{\nu}}({\color{blue}x})$&  $\leftrightsquigarrow$  & $KK$-wave (0-brane) \\
level 1&  $A_{\hat{\mu}\hat{\nu} \hat{\rho}}({\color{blue}x})$ + $g_{\hat{\mu} \hat{\nu}}({\color{blue}x})$&  $\leftrightsquigarrow$  & $M2$ (2-brane) \\
level 2&  $A_{\hat{\mu}_1...\hat{\mu}_6}({\color{blue}x})$ + $g_{\hat{\mu} \hat{\nu}}({\color{blue}x})$&  $\leftrightsquigarrow$  & $M5$ (5-brane) \\ 
level 3&  $A_{\hat{\mu}_1...\hat{\mu}_8,\hat{\nu}}({\color{blue}x})$ + $g_{\hat{\mu} \hat{\nu}}({\color{blue}x})$ &  $\leftrightsquigarrow$  & $KK6$-monopole 
\end{tabular}
}
\caption {Link between the lowest level up to level $3$ of the brane $E_8^{++}$-invariant theory and the branes of M-theory.}
\end{table}

 Moreover, intersections rules are neatly encoded in $\gdp_{B}$ by orthogonality conditions on the positive real roots characterizing each branes. 
\section{Weyl transformations and their consequences}
In this section, we review some consequences of Weyl transformations. First, a Weyl transformation on a generator $T$ of $\G$ can be expressed as a conjugaison by a group element $U_W$ of $\G$: $T \longrightarrow  U_W T U_W^{-1}$. Because of the non-commutativity of Weyl transformation with the temporal involution $\Omega$:
\begin{eqnarray}
U_W \Omega T U_W^{-1}=  \Omega'  U_W T U_W^{-1} \, ,
\end{eqnarray}
different Lorentz signatures $(t,s)$ (where $ t \,(s)$ is the number of time (space) coordinates) can be obtained \cite{Englert2004ph}.  More precisely, Weyl reflections of the gravity line do not change the global Lorentz signature $(t,s)$ but it change only the identification of time coordinates. In fact, only Weyl reflections with respect to roots not belonging to the gravity line can change the global signature of the theory.\\

For instance the signatures found for $E_8^{+++}$ are $(1,10)$, $(2,9)$, $(5,6)$, $(6,5)$ and $(9,2)$ \cite{Keurentjes2004bv}. These signatures match perfectly with the signatures changing dualities. Indeed, we can interpret the Weyl transformation with respect to root $\alpha_{11}$ (not belonging to gravity line) as a double T-duality in the direction $9$ and $10$ with an exchange of these directions \cite{Obers1998rn, Englert2003zs}. Moreover, these signatures match also with the exotic phases of M-theories (M' and M*) \cite{Hull1998ym}.\\

The previous construction has been generalized and the signatures for all $\G$ have been found \cite{deBuyl2005it,Keurentjes2005jw}. 
\section{Conclusions and perspectives}
Many properties of space time are neatly encoded in Kac-Moody algebras (branes, intersection rules, T-duality, \ldots). So the study of $\gdp$ and $\G$ constitutes an interesting approach to understand the gravitational theories coupled to matter which is conceptually different from the Einstein approach. We have now to answer to the fundamental question of all this approach: are these symmetries only a consequence of the compactification process or are they effectively symmetries of the uncompactified theory? This question is related to the significance of the infinite number of fields. Progress is made in this direction \cite{prog}.
%%%%%%%%%%%%%%%%%%%%%%%%%%%%%%%%%%%%%%%%%%%%%%%%%%%
\section*{Acknowledgments}
 I cordially thank Laurent Houart for his encouragement and his stimulation. I also thank Sophie and Pierre de Buyl, Nathan Goldman for their technical support and Vincent Wens for his reading of the paper.

\end{document}